\documentclass[aps,twocolumn,groupedaddress,superscriptaddress,floatfix,prb]{revtex4}
\usepackage{amsmath}
\usepackage{amssymb}
\usepackage{graphicx}
\usepackage{textcomp}
\usepackage{bm}	

\usepackage{booktabs}
\usepackage{siunitx}

\newcommand{\Ss}{\mathcal{S}}
\newcommand{\A}{\mathcal{A}}
\newcommand{\PP}{\mathcal{P}}
\newcommand{\ket}[1]{| #1 \rangle}
\newcommand{\bra}[1]{\langle #1 |}

\usepackage[usenames,dvipsnames]{color}	

\begin{document}

\title{Steady state entanglement beyond thermal limits}

\author{F. Tacchino}
\affiliation{Dipartimento di Fisica, Universit\`a di Pavia, via Bassi 6, I-27100, Pavia, Italy}
\author{A. Auff\`eves} \email{alexia.auffeves@neel.cnrs.fr}
\affiliation{CNRS and Universit\'{e} Grenoble Alpes, Institut N\'{e}el, F-38042, Grenoble, France}
\author{M. F. Santos}
\affiliation{Instituto de F\'{i}sica, Universidade Federal do Rio de Janeiro, CP68528, Rio de Janeiro, RJ 21941-972, Brazil}
\author{D. Gerace}\email{dario.gerace@unipv.it}    
\affiliation{Dipartimento di Fisica, Universit\`a di Pavia, via Bassi 6, I-27100, Pavia, Italy}

\begin{abstract} 
Classical engines turn thermal resources into work, which is maximized for reversible operations.
The quantum realm has expanded the range of useful operations beyond energy conversion, and
incoherent resources beyond thermal reservoirs. This is the case of entanglement generation in
a driven-dissipative protocol, which we hereby analyze as a continuous quantum machine. We
show that for such machines the more irreversible the process the larger the concurrence. Maximal
concurrence and entropy production are reached for the hot reservoir being at negative effective 
temperature, beating the limits set by classic thermal operations on an equivalent system.
\end{abstract}

\maketitle

\textit{Introduction}.
{Engines rely on the ability to perform useful operations by exploiting incoherent resources. Typical examples are classical thermal machines, which extract work from a ``working fluid'' upon transfer of heat from a hot to a cold bath: the efficiency of such  machines is defined by the ratio between the work produced and the heat absorbed from the hot bath. In the quantum realm, the working medium may provide a non-classical inner structure. New out-of-equilibrium {scenarii} can thus be envisioned, in which different quantized transitions are coupled to independent heat baths. This strategy can be used, \textit{e.g.},  {to invert the population of some medium by optical pumping}, and to extract work by stimulating the transition: as a matter of fact, lasers and micro-masers have for long been interpreted as out-of-equilibrium heat engines \cite{Scovil1959, Scully01}.}

{More generally, {optical pumping schemes are} used to selectively prepare {and maintain} the working medium in a given non-trivial target steady state {that is different from its thermal equilibrium state}. In this spirit, the potential of achieving steady state entanglement of pairs of qubits through quantum optical bath engineering has started to be explored \cite{Braun2002,Schneider2002,Kim2002,Benatti2002,Lambert2007,DelValle2007,
Kraus2008,Diehl2008,Li2009,Verstraete2009,DelValle2011,Reiter2013,Aron2014,Vasco2016}, most of the attention being {focused on} thermal baths as purely incoherent sources of non classical correlations \cite{Camalet2011,Carvalho2011,Bellomo2013,Bellomo2015,BohrBrask2015,Tavakoli2017}; however, for the latter protocols the amount of entanglement that can be generated without any additional feedback or filtering operation is typically rather modest \cite{BohrBrask2015,Tavakoli2017}. {By using reservoirs acting on collective degrees of freedom of the qubits, the upper theoretical limit for the concurrence can be increased to} ${C} = 1/3$, which is asymptotically reached only under unrealistically large temperature gradients between the two reservoirs \cite{Camalet2011,Bellomo2013}.}

{While no work is effectively extracted, these operations are still typical of a machine, for they reach a useful goal (\textit{i.e.}, the preparation of some out-of-equilibrium desired steady state) by exploiting incoherent resources. This calls for the definition of new criteria to assess the performance of such devices operating in the continuous regime \cite{Kosloff2014,Uzdin2015,Goold2016}. On a parallel route, looking at the inner structure of a quantum system as a thermodynamic resource can be exploited, e.g., to increase the efficiency of  miniaturized engines \cite{Scully2003,Rossnagel2014,Klaers2017}. In this Letter, we {study and thermodynamically characterize an optical pumping-based quantum machine, which allows for the generation of steady state entanglement in a bipartite system coupled to two incoherent reservoirs at different temperatures. We show that the machine performs all the better as its lead to larger amounts of steady-state entropy production, consistently with the irreversible character of the protocol. Our study is based on a general definition of entropy production taken from stochastic thermodynamics \cite{Seifert08,Crooks08,Horowitz13,Manzano15,Elouard17}, which interestingly allows to extend the theoretical analysis to the case of reservoirs with negative effective temperatures. This generalized definition of baths allows to increase the amount of steady state entanglement beyond the known limits imposed by classical heat baths at thermal equilibrium \cite{Camalet2011,Bellomo2013}. Finally, we propose a practical realization of the quantum thermal machine based on two independent and incoherently pumped qubits that are coupled to a leaky cavity mode \cite{Temnov2009,Auffeves2011}. }

\textit{Optical pumping and steady state entanglement}. 
{As an elementary model of a driven-dissipative quantum machine producing steady state entanglement, we consider a generic bipartite quantum system consisting of two independent qubits of ground and excited levels respectively denoted $\ket{0}_i$ and $\ket{1}_i$ ($i=1,2$). The internal level structure of the composite system is then characterized by a diamond-like scheme, 
as represented in Fig.~\ref{fig1}a, with degenerate transitions energies $\omega_0 = \omega_A - \omega_G=\omega_S - \omega_G$, where
$\sqrt{2}  |S\rangle = (|0\rangle_1  |1\rangle_2 + |1\rangle_1 |0\rangle_2)$ and $ \sqrt{2}  |A\rangle =(|0\rangle_1  |1\rangle_2 - |1\rangle_1  |0\rangle_2)$ are the two maximally entangled Bell states, respectively. 
The goal is to generate steady state entanglement by optically pumping the system in one of these states with high probability and by using incoherent resources, which can practically be realized by engineering some unbalance between the steady state population of $\ket{S}$ and $\ket{A}$.}

\begin{figure*}
\centering
\includegraphics[width=0.86\textwidth]{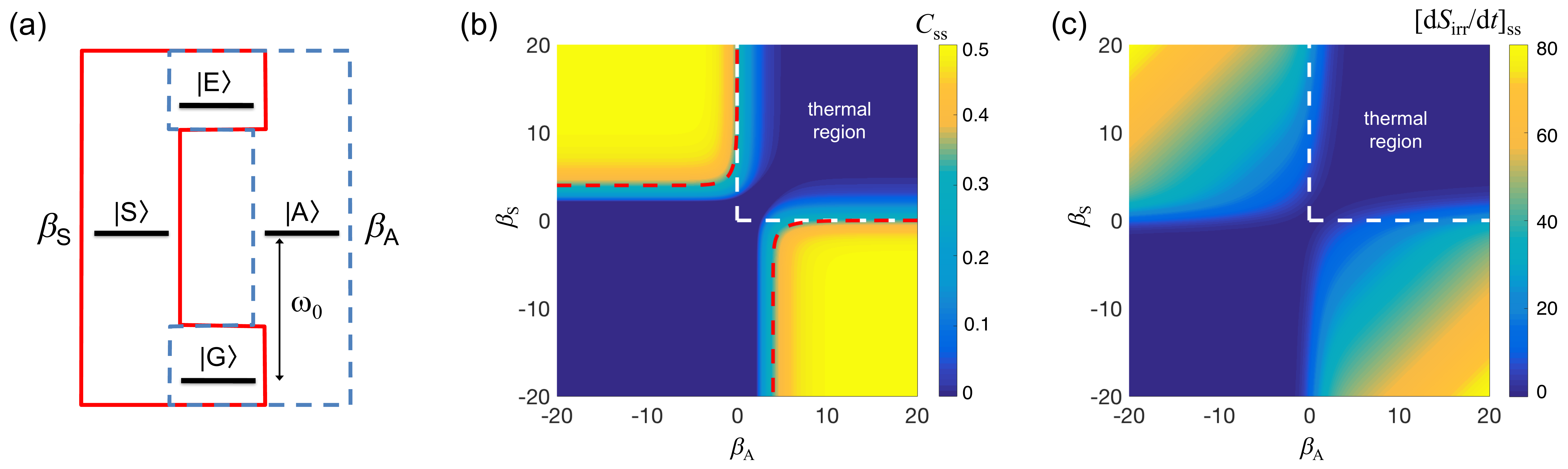}
\caption{
(a) Elementary model of a driven-dissipative quantum thermal machine with diamond-like internal level structure and degenerate symmetric ($S$) and antisymmetric ($A$) states;
the collective eigenstates are assumed to be connected to two independent reservoirs, defined through their effective temperatures $\mathcal{T}_S=\omega_0 /\beta_S$ and $\mathcal{T}_A=\omega_0 /\beta_A $, respectively.
(b) Steady state concurrence, Eq. (\ref{eq:ss_concurrence}), plotted against $\beta_A$ and $\beta_S$, respectively. 
White dashed lines mark the thermal region ($\beta_{A,S}\geq 0$); the red dashed curves show the contour line for the limiting value $C=1/3$.
(c) Rate of entropy production in steady state, Eq. (\ref{eq:entropy_rate}), plotted in the same $(\beta_A,\beta_S)$ plane as panel {(b)} and normalized to $\omega_0$ and $\Gamma^+$.
}\label{fig1}
\end{figure*}

We assume the two qubits to be coupled to two independent effective baths, each one acting on a collective degree of freedom. The inverse temperatures of these baths are denoted as $\beta_A$ and $\beta_S$, which are allowed to assume negative values \cite{Kittel1980}. The dynamics of the open quantum system is completely described by the master equation \cite{Breuer2002} for the density matrix ($\hbar=1$ and $k_{B}=1$ in the following)
\begin{equation}\label{eq:mastereq}
\partial_t \rho = \, i [ {\rho}, \hat{H}_0 ] +  \mathcal{L}(\rho)
\end{equation}
where $\hat{H}_0 = \omega_0 \, (\hat{c}^{\dagger}_1 \hat{c}_1 + \hat{c}^{\dagger}_2 \hat{c}_2 )$ is the Hamiltonian, and 
\begin{equation}\label{eq:effective_collective_baths}
\mathcal{L}(\rho) = \sum_{i=A,S} \left[ \frac{\Gamma^{+}_i}{2} \mathcal{D}_{\hat{J}^{\dagger}_i}(\rho) + \frac{\Gamma^{-}_i}{2}\mathcal{D}_{\hat{J}_i} (\rho) \right]
\end{equation}
is the Liouvillian operator in Lindblad form, with $\mathcal{D}_{\hat{o}}(\rho) = 2\hat{o}\rho \hat{o}^\dagger - \{\hat{o}^\dagger \hat{o}, \rho\}$. 
{We have introduced the collective operators $\hat{J}_S=\hat{c}_1 + \hat{c}_2 $ and  $\hat{J}_A=\hat{c}_1 - \hat{c}_2 $ respectively, where  $\hat{c}_i$ ($\hat{c}_i^{\dagger}$, $i=1,2$)  are destruction (creation) operators obeying anticommutation rules $\{ \hat{c}_i, \hat{c}_j^{\dagger} \} = \delta_{ij}$.} We assume a common pumping rate for the two collective modes, \textit{i.e.} $\Gamma^{+}_{i}=\Gamma^{+}$ for $i=A,S$, while we will allow for independent dissipation rates, $\Gamma^{-}_{S}$ and $\Gamma^{-}_A$, verifying
\begin{equation}\label{eq:temperatures}
\frac{\Gamma^{+}}{\Gamma^{-}_{S}}= e^{- \beta_S } \,\,\,\,  ; \,\,\,\,\, \frac{\Gamma^{+}}{\Gamma^{-}_{A}}= e^{- \beta_A }
\end{equation}
in which the effective temperatures of the baths are given in units of $\omega_0$. \\
{The amount of steady state entanglement that can be generated is quantified from the degree of non-separability of the given steady state $\rho_{SS}$, \textit{i.e.} the solution of the linear equation $ [ {\rho}_{SS}, \hat{H}_0 ] = i \mathcal{L}(\rho_{SS})$. As an entanglement measure we hereby use the  concurrence \cite{Wootters1998,Wootters2001}, an entanglement monotone function defined for bipartite quantum systems as $C(\rho) = \max \{ 0, \lambda_1 - \lambda_2 - \lambda_3 - \lambda_4\}$, where $\lambda_i^2$ are the eigenvalues of the Hermitian matrix $\rho\tilde{\rho}$ ordered as 
$\lambda_1\geq \lambda_2 \geq \lambda_3 \geq \lambda_4$, and $\tilde{\rho} = (\sigma_y \otimes \sigma_y)\rho^*(\sigma_y \otimes \sigma_y)$. 
For maximally entangled pure states such as, \textit{e.g.}, Bell states, the concurrence is bound to $C[\rho] = 1$.
For the model above, the steady state concurrence can be analytically solved as (see App.~A)
\begin{equation}\label{eq:ss_concurrence}
C_{SS} \equiv C(\rho_{SS}) = \max \left\{0,({N_1}-{N_2})/{d}\right\}
\end{equation}
with
\begin{equation}
\begin{split}
N_1 = &\, \left|\A\left({\Ss}/{2}-1\right)\right| \\
N_2 = &\, \left[\left({\Ss}/{2}+1\right)\left(2\Ss^2+2\PP(\Ss-2)\right)\right]^{1/2} \\
d = &\, 1 + \Ss^2 + {\Ss}(\PP +3)/2 - \PP
\end{split}
\end{equation}
in which we defined $\Ss=\exp(\beta_A) + \exp(\beta_S)$, $\A=\exp(\beta_A) - \exp(\beta_S)$, and $\PP=\exp(\beta_A+\beta_S)$.

A plot of Eq. \eqref{eq:ss_concurrence} is given in Fig. \ref{fig1}b as a function of the two inverse temperatures. 
As expected, the steady state is fully separable ($C_{SS}=0$) for balanced reservoirs, \textit{i.e.} when $\beta_S \simeq \beta_A$.
In this case, most of the stationary population is either in $|G\rangle$ or $|E\rangle$ on average, while the rest is in an equal mixture of the two Bell states $|A\rangle$ and $|S\rangle$.
On the other hand, an unbalance in the two thermal reservoirs allows for driving the system in a non-separable steady state, with the population of either $|A\rangle$ or $|S\rangle$ dominating over the other. 
The amount of entanglement is limited to the value $C=1/3$ when classical thermal reservoirs at positive temperatures are assumed (see dashed lines superimposed to the color scale plot), as also inferred from the analytic expression above (App.~A). This limiting value is reached when the cold bath is at zero temperature goes while the hot one is at infinite temperature. In such a case the population of the Bell states is unbalanced, such that $1/3$ of the weight is in the entangled state coupled to the hot bath, while the rest of the, i.e. $2/3$, is in the ground state, $|G\rangle$. A similar  result was  found in alternative models of bipartite quantum systems coupled to thermal reservoirs \cite{Camalet2011,Bellomo2013}. 

Going beyond previous studies, we see that the optical pumping can be improved by relaxing the conditions of real thermal reservoirs with positive temperatures. If negative temperatures are authorized, maximal steady state entanglement is reached when the hot bath is at effective negative temperature, while the cold one at a positive and small one (Fig. \ref{fig1}b). Then, the system is pumped into the maximally entangled state coupled to the negative effective temperature bath with $1/2$ stationary probability, with zero probability in the other, giving the limiting value $C_{SS}=1/2$. This is a key result of this work: a bipartite quantum system can be optically pumped into a maximally entangled steady state by exploiting purely incoherent resources, with the largest concurrence reaching the limiting value of 0.5 if one of the two Bell states is coupled to a bath at negative effective temperature. In the absence of feedback or further purification of the steady state \cite{Wang2005,Carvalho2007,Tavakoli2017}, this is the theoretical limiting value. Notice that the regions with the highest concurrence are all outside the thermal region, which could in principle be reached with classic thermal baths. Notice also that the lower left region, corresponding to both reservoirs being at negative effective temperature, gives $C_{SS}=0$ due to the largest occupancy of the fully separable $|E\rangle$ state, \textit{i.e.} corresponding to the population inversion of the diamond at large pumping.

\textit{Entropy production and irreversibility}. 
{The optically pumped bipartite system is now analyzed in terms of its thermodynamic properties. The whole protocol aims at driving and maintaining a quantum system out of equilibrium and therefore, is irreversible by nature. The degree of irreversibility is quantified by the rate of steady state entropy production, $\dot{S}_\mathrm{irr}=[\mathrm{d}{S}_\mathrm{irr}/\mathrm{d}t]_{SS}$. If the reservoirs are real thermal baths, this rate is classically given by}
\begin{equation} \label{eq:entropy_rate}
\begin{split}
\dot{S}_{\mathrm{irr}}[\rho_{SS}] = -\beta_A \dot{Q}_A[\rho_{SS}] - \beta_S \dot{Q}_S[\rho_{SS}] \geq 0 \, \, ,
\end{split}
\end{equation}
where the steady state heat currents are defined as $\dot{Q}_i[\rho_{SS}]= \mathrm{Tr}\{ H_0 \mathcal{L}_i (\rho_{SS}) \}$, {with $\mathcal{L}_i (\rho_{SS})$ ($i=A,S$) as in Eq.\ \eqref{eq:effective_collective_baths}}, which {verify} $\dot{Q}_A[\rho_{SS}] + \dot{Q}_S[\rho_{SS}] = 0$. In this classical case, Eq.\ \eqref{eq:entropy_rate} simply corresponds to the increase of entropy of the isolated system consisting of the two qubits and the two baths, consistently with the Second Law.
Stochastic thermodynamics allows to extend the concept of entropy production to new regimes where incoherent resources do not reduce to thermal baths \cite{Seifert08,Crooks08,Horowitz13,Manzano15,Elouard17}. Here entropy production is defined at the single realization level, by comparing the respective probabilities of the realization in the direct and in some fictitious, reversed protocol. Such definition verifies the Second Law (the rate of entropy production is positive on average), and matches Eq.\ \eqref{eq:entropy_rate} if the reservoirs are thermal. Remarkably, based on stochastic thermodynamics it can be shown that the validity of Eq.\ \eqref{eq:entropy_rate} still holds in the case of reservoirs at negative effective temperature (see App.~B). The results are shown in Fig. \ref{fig1}c as a function of $\beta_A$ and $\beta_S$, displaying a striking correlation with the concurrence plot: our protocol can be seen as a machine operating in the steady-state regime, whose ability to generate entanglement is maximized with the entropy production rate.

\textit{A cavity QED-based implementation}.
A natural question is whether the theoretical model in Eq.\ \eqref{eq:effective_collective_baths} can be practically realized in a physical system that is amenable to experimental implementation. We show here that this is the case for a quite straightforward cavity QED situation in which two independent and incoherently pumped qubits are coupled to a single radiation mode of an electromagnetic resonator, as schematically represented in Fig.~\ref{fig2}a. 
Specifically, we consider a pair of point-like two level systems that are resonantly ($\omega_{\mathrm{cav}}=\omega_0$) coupled to a single-mode resonator at the same rate $g\ll\omega_0$, such that rotating wave approximation is justified and their Hamiltonian is a two-emitters Tavis-Cummings model
\begin{equation}\label{Hamrot}
\begin{aligned}
\hat{H}_{TC} = & \sum_{i=1}^2 {\omega}_0 \, \hat{c}^{\dagger}_i \hat{c}_i + \omega_{\mathrm{cav}} \, \hat{a}^\dagger \hat{a} + \sum_{i=1}^2 g \, (\hat{c}^{\dagger}_i\hat{a} + \hat{c}_i \hat{a}^\dagger)
\end{aligned}
\end{equation}
where $\hat{a}$ ($\hat{a}^\dagger$) is the destruction (creation) operator of the single-mode cavity photons.
The master equation describing the driven-dissipative system of Fig.~\ref{fig1}b is thus $\partial_t \rho = \, i [ {\rho}, \hat{H}_{TC} ] +  \mathcal{L}(\rho)$, where the full Liouvillian explicitly reads
\begin{equation}\label{eq:liouvillian}
\mathcal{L}(\rho) = \frac{p}{2}  \sum_{i=1,2} \mathcal{D}_{\hat{c}^{\dagger}_i}(\rho) + 
\frac{\gamma}{2} \sum_{i=1,2} \mathcal{D}_{\hat{c}_i}(\rho)   +  \frac{\kappa}{2} \mathcal{D}_{\hat{a}}(\rho) 
\end{equation}
in which $p$ and $\gamma$ are the incoherent pumping and relaxation rates of the two (identical) qubits, and $\kappa$ describes the photon emission rate from the cavity.
Notice that an incoherent pumping scheme is realized whenever high-energy excitations relax to a well defined ground state transition at a certain rate, even if the original source of excitation can be a coherent one (\textit{e.g.}, an off-resonant laser). 
The steady state of the full model can be solved numerically (see App.~C). This model has been previously analyzed, \textit{e.g.}, in Ref. \onlinecite{Auffeves2011}, when it was evidenced that a steady state subradiant regime exists over a broad range of values $\kappa/g$, under weak pumping conditions $p \ll g$. This system can be effectively described in the collective spin basis by adiabatically eliminating the cavity mode \cite{Temnov2009}, which is fully justified for $\kappa/g>1$. In fact, under such conditions the cavity only acts as an additional dissipation channel in the reduced two-qubits subspace \cite{Auffeves2011} in which each qubit  is further relaxed at a rate $\Gamma=4g^2 / \kappa$ in addition to the intrinsic spontaneous emission at rate $\gamma$. Hence Eq.\ \eqref{eq:liouvillian} can be recast exactly as Eqs.\ \eqref{eq:mastereq} and \eqref{eq:effective_collective_baths}, after straightforward algebra with the following relations
\begin{equation}\label{eq:parameters}
\Gamma^+ = p/2 \,\,\,; \,\,\,  \Gamma^{-}_A=\gamma/2 \,\,\,; \,\,\, \Gamma^{-}_S=\Gamma + \gamma/2
\end{equation}
The effective temperatures result now from combinations of the physical parameters of the model: $\beta_A = \log \left[\gamma/p_x\right]$ and $\beta_S = \log \left[(\gamma + 2\Gamma)/p_x\right]$.

\begin{figure}[t]
\begin{center}
\includegraphics[width=0.42\textwidth]{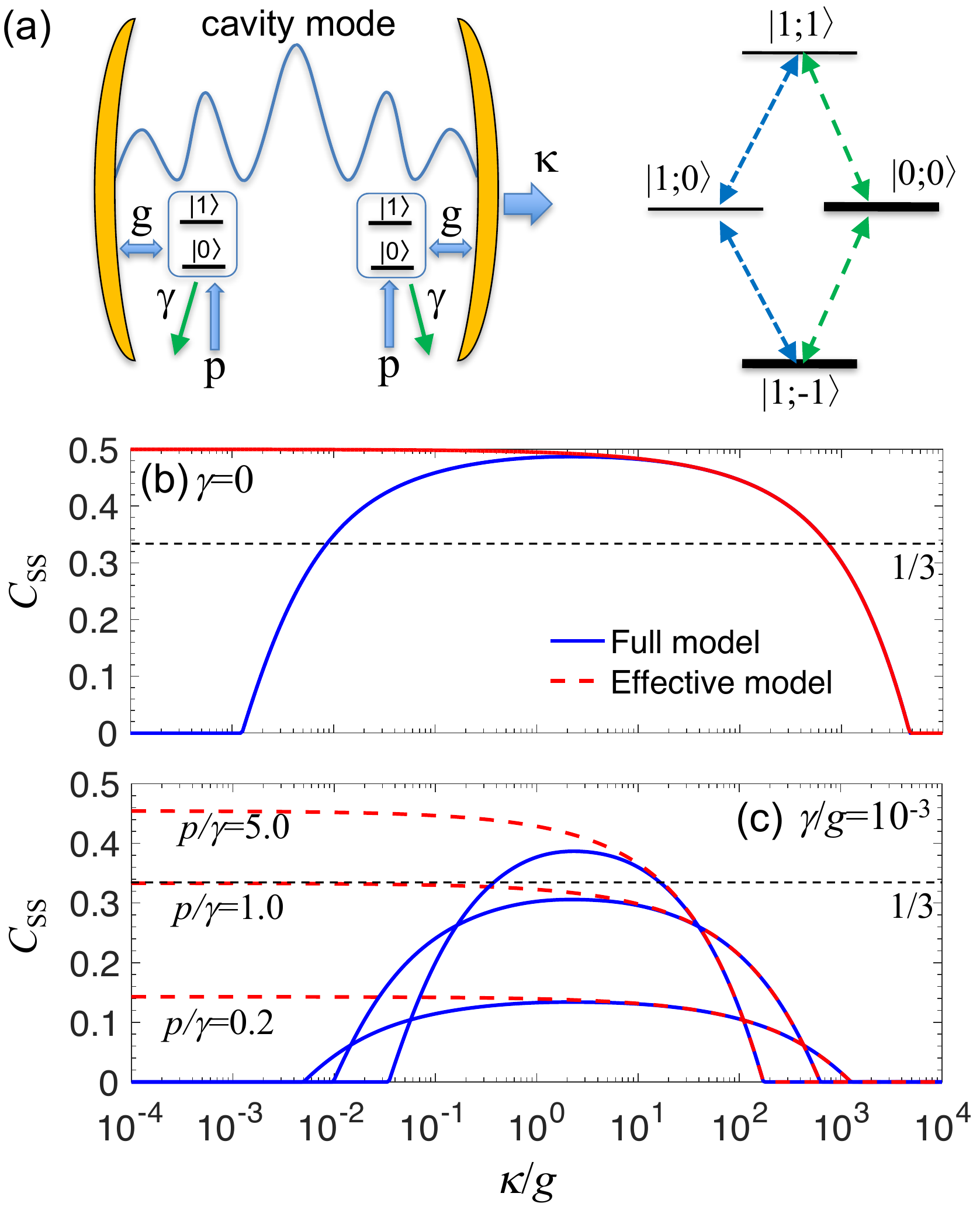}
\caption{ 
(a) Quantum optical scheme of the full cavity QED model model representing a pair of two-level emitters is coupled to the same lossy cavity mode and incoherently pumped by an external drive, and the corresponding
level scheme of the effective model after adiabatic elimination of the cavity degree of freedom.  
(b) Comparison between the steady state concurrence calculated numerically for the full quantum optical model, Eq. (\ref{eq:liouvillian}), and analytically for the effective model, Eq. (\ref{eq:effective_collective_baths}) with parameters as in Eqs. (\ref{eq:parameters}), 
as a function of the cavity dissipation rate for the ideal case of negligible qubits relaxation rate ($\gamma=0$, at pumping strength $p=0.0002g$);  (c) same comparison for finite $\gamma$ and varying pumping strength: $p=0.0002g$, $p=0.001g$, $p=0.005g$. Here, the qubit-cavity coupling rate is chosen to be $g/\omega_0 = 0.001$.
}\label{fig2}
\end{center}
\end{figure}

In the subradiant regime the system is optically pumped in the dark $|A\rangle=|0;0\rangle$ state (\textit{i.e.}, the singlet in the $|J; M_J\rangle$ notation for eigenstates of the total angular momentum), thus creating an imbalanced population with respect to the $|S\rangle=|1;0\rangle$ (triplet) state \cite{Auffeves2011}, as schematically represented in the diamond-like level structure of Fig. \ref{fig2}a. 
This is confirmed by plotting the steady state concurrence of the two qubits in Fig. \ref{fig2}b,c, which is evidently different from zero only when $\kappa/g$ falls in the subradiant sector of the model. There exists an optical pumping range for which the system reaches its maximal concurrence, which also depends on $\gamma/g$. In particular, the maximal value is $C_{SS}\simeq 0.4$ around $p/ \gamma \simeq 5$ for the case shown in Fig.~\ref{fig2}, but it can be even larger and approaching the $C_{SS}=0.5$ limit for smaller values of $\gamma/g$ (see, \textit{e.g.}, full numerical results in App.~C).  \\
First, in Fig. \ref{fig2}b we show the ideal result for $\gamma=0$, corresponding to the negative effective temperature reservoir coupled to the dark state, which gives the limiting value $C_{SS}\to 0.5$ when $p/\Gamma \to 0$ (in agreement with the results in Fig. \ref{fig1}); the full model only follows the effective model for $\kappa/g >1$, \textit{i.e.} until the adiabatic elimination of the cavity mode holds.
At difference with the general model of the previous section, here the $A$-$S$ symmetry is broken since only the antisymmetric state is dark and, from Eq.\ \eqref{eq:parameters}, $\beta_S > \beta_A$. Hence, with reference to Fig.\ \ref{fig1}, only the part above the $\beta_S = \beta_A$ diagonal should be considered when dealing with this cavity QED implementation. 
In Fig. \ref{fig2}c we show the behavior of the steady state concurrence for $\gamma=10^{-3} g$, which is usually the case in most practical realizations of this quantum optical model, e.g. in solid-state cavity QED.  While it is evident that the regime of non-separability narrows in $\kappa$ as $p$ increases, it should also be noted that for the proper values of $\kappa$ the thermodynamic limit is overcome (\textit{i.e.}, $C_{SS} > 1/3$) as soon as $p> \gamma$. The latter condition corresponds to the onset of negative effective temperature for the dark state reservoir (App.~C).\\

\textit{Discussion}.
{We propose and thermodynamically analyze a new protocol to generate steady state entanglement of a bipartite quantum system from incoherent resources. We show that this scheme can be interpreted as a continuous quantum engine whose performances are optimized when the entropy production rate is maximal. This effect makes this class of engines very different from classical engines, whose yield is usually maximized in the reversible regime. }
We finally highlight the potential interest of this results from an experimental point of view: there are different platforms where these results at the forefront between quantum optics, quantum information, and quantum thermodynamics could be tested, ranging from semiconductor quantum dots spatially and spectrally matched to photonic nanoresonators \cite{Hennessy2007,Dousse2008,Lyasota2015}, to superconducting circuit quantum electrodynamics devices \cite{Shankar2013,Kimchi2016,DiCarlo2009}. More quantitative information is provided in App.~E.

\textit{Acknowledgements}. 
The authors acknowledge useful discussions and suggestions from M. Campisi, S. Carretta, E. Mascarenhas, M. Lostaglio, M. Richard, A. Tavakoli, F. Troiani, H. E. T\"{u}reci, J. P. Vasco Cano. 
This work was partly supported from COST Action MP1403 ``Nanoscale Quantum Optics'' through the Short Term Scientific Mission (STSM) program, 
the Italian Ministry of Education and Research (MIUR) through PRIN Project 2015 HYFSRT ``Quantum Coherence in Nanostructures of Molecular Spin Qubits'', 
the Brazilian funding agency CNPq through project  No. 305384/2015-5 and the PVE-Ci\^{e}ncia Sem Fronteiras Project No. 407167/2013-7, 
the CNRS French-Brazilian PICS program ``Thermodynamics of Quantum Optics''.

\appendix

\section{Steady state concurrence of the bipartite quantum system}
From the master equation
\begin{equation}\label{eq:mastereq}
\partial_t \rho = \, i [ {\rho}, \hat{H}_0 ] +  \mathcal{L}(\rho)
\end{equation}
where $\hat{H}_0 = \omega_0 \, (\hat{c}^{\dagger}_1 \hat{c}_1 + \hat{c}^{\dagger}_2 \hat{c}_2 )$
and 
\begin{equation}\label{eq:effective_collective_baths}
\mathcal{L}(\rho) = \sum_{i=A,S} \left[ \frac{\Gamma^{+}_i}{2} \mathcal{D}_{\hat{J}^{\dagger}_i}(\rho) + \frac{\Gamma^{-}_i}{2}\mathcal{D}_{\hat{J}_i} (\rho) \right]
\end{equation}
one can find the steady state $\rho_{SS}$ by imposing the condition
\begin{equation}
\partial_t \rho = 0
\end{equation}
The solution expressed in the computational basis $\{|00\rangle, |01\rangle, |10\rangle, |11\rangle\}$ has the general form
\begin{equation}
\rho_{SS} =
\begin{pmatrix}
\rho_{00} & 0 & 0 & 0 \\ 
0 & \rho_{01} & \rho_{c} & 0 \\ 
0 & \rho_{c}^* & \rho_{10} & 0 \\ 
0 & 0 & 0 & \rho_{11}
\end{pmatrix} 
\label{eq:rho_ss}
\end{equation}
This can be easily interpreted as a consequence of the fact that the dissipative part of the Liouvillian involves coherent superpositions only in the $\{|S\rangle,|A\rangle\}$ subspace. 
According to the formal definition \cite{Wootters1998,Wootters2001}, the concurrence of a two-qubit density matrix $\rho$ can be computed as follows:
\begin{itemize}
\item define $\tilde{\rho} = (\sigma_y \otimes \sigma_y)\rho^*(\sigma_y \otimes \sigma_y)$
\item find the spectral decomposition of $\rho\tilde{\rho}$ as
\begin{equation}
\rho\tilde{\rho} = \sum_{i=1}^4 \lambda_i^2 |\psi_i\rangle\langle \psi_i|
\end{equation}
\item after ordering the eigenvalues as 
\begin{equation}
\lambda_1\geq \lambda_2 \geq \lambda_3 \geq \lambda_4
\end{equation}
the concurrence of $\rho$ is
\begin{equation}
C(\rho) = \max \{\lambda_1 - \lambda_2 - \lambda_3 - \lambda_4, 0\}
\end{equation}
\end{itemize}
For $\rho = \rho_{SS}$ as given in \eqref{eq:rho_ss} this reduces to
\begin{equation}
C_{SS} \equiv C(\rho_{SS}) = 2\max\left\{0, |\rho_{c}|-\sqrt{\rho_{00}\rho_{11}}\right\}
\end{equation}
and, using the explicit solution
\begin{equation}
C(\rho_{SS}) = \max \left\{0,({N_1}-{N_2})/{d}\right\}
\end{equation}
with
\begin{equation}
\begin{split}
N_1 = &\, \left|(\Gamma^{-}_1/\Gamma^{+} - \Gamma^{-}_2/\Gamma^{+})\left(\Gamma^{-}_1/\Gamma^{+} + \Gamma^{-}_2/\Gamma^{+}-2\right)\right| \\
N_2 = &\, \Big[\left(\Gamma^{-}_1/\Gamma^{+}+\Gamma^{-}_2/\Gamma^{+}+2\right)\Big((\Gamma^{-}_1/\Gamma^{+})^2(\Gamma^{-}_2/\Gamma^{+}) \\  & \, + (\Gamma^{-}_1/\Gamma^{+})^2 + (\Gamma^{-}_2/\Gamma^{+})^2 + (\Gamma^{-}_2/\Gamma^{+})^2(\Gamma^{-}_1/\Gamma^{+})\Big)\Big]^{1/2} \\
d = &\, \Big(1+(\Gamma^{-}_1/\Gamma^{+})^2 + (\Gamma^{-}_2/\Gamma^{+})^2 \\ & \, + (\Gamma^{-}_1/\Gamma^{+})(\Gamma^{-}_2/\Gamma^{+}) + (3/2)(\Gamma^{-}_1/\Gamma^{+} + \Gamma^{-}_2/\Gamma^{+}) \\ & \, + (1/2)(\Gamma^{-}_1/\Gamma^{+} + \Gamma^{-}_2/\Gamma^{+})(\Gamma^{-}_1/\Gamma^{+})(\Gamma^{-}_2/\Gamma^{+})\Big)
\end{split}
\end{equation}
By using the definitions of effective temperatures (in units of $\omega_0$) as given in the paper
\begin{equation}\label{eq:temperatures}
\frac{\Gamma^{+}}{\Gamma^{-}_{S}}= e^{- \beta_S} \,\,\,\,  ; \,\,\,\,\, \frac{\Gamma^{+}}{\Gamma^{-}_{A}}= e^{- \beta_A}
\end{equation}
one gets, after some algebra
\begin{equation}
\begin{split}
N_1 = &\, \left|\A\left({\Ss}/{2}-1\right)\right| \\
N_2 = &\, \left[\left({\Ss}/{2}+1\right)\left(2\Ss^2+2\PP(\Ss-2)\right)\right]^{1/2} \\
d = &\, 1 + \Ss^2 + {\Ss}(\PP +3)/2 - \PP
\end{split}
\end{equation}
in which we defined $\Ss=\exp(\beta_A) + \exp(\beta_S)$, $\A=\exp(\beta_A) - \exp(\beta_S)$, and $\PP=\exp(\beta_A+\beta_S)$. It is easy to realize that both effective temperatures are treated symmetrically in the final expression, meaning that the system can in principle rely both on the symmetric or antisymmetric maximally entangled state to produce non-zero concurrence.

Finally, it is also easy to show explicitly the behavior of the concurrence in some instructive cases. First, let us put $\beta_A = \beta_S$: this corresponds to equal effective temperatures that always produce a separable steady state. Indeed, we have $\A = 0 \Rightarrow N_1 = 0$ and $N_2 > 0$, implying $C(\rho_{SS}) = 0$. One can also show that under these conditions the collective dissipators appearing in the master equation decouple into local ones. On the other hand, when $\beta_A = 0$ and $\beta_S \to +\infty$ we have $N_1 \simeq \exp(2\beta_S)/2$, $N_2 \simeq \exp((3/2)\beta_S)$ and $d \simeq (3/2)\exp(2\beta_S)$, and we reach the thermal limit
\begin{equation}
C(\rho_{SS}) \simeq \frac{2}{3}\left(\frac{1}{2} - e^{-\beta_S/2}\right) \to \frac{1}{3}
\end{equation}
Finally, we can consider the extreme case $\beta_A \to - \infty$, for which $\Ss \simeq \exp(\beta_S)$, $\A \simeq - \exp(\beta_S)$ and $\PP \simeq 0$ for every finite value of $\beta_S$. The factors appearing in the formula for the concurrence are now $N_1 \simeq \exp(\beta_S)[\exp(\beta_S)/2 -1]$, $N_2 \simeq \exp(\beta_S)[\exp(\beta_S)+2]^{1/2}$ and $d \simeq 1 + \exp(2\beta_S) +(3/2)\exp(\beta_S)$, and if $\beta_S$ is positive and large enough (to be precise, we should still ask $|\beta_A|\gg \beta_S$ so that \textit{e.g.} the $\PP \simeq 0$ limit holds) we get the maximum
\begin{equation}
C(\rho_{SS}) \simeq \frac{(1/2)e^{2\beta_S}-e^{\beta_S}}{1+e^{2\beta_S} +(3/2)e^{\beta_S}} - \frac{e^{\beta_S}(e^{\beta_S}+2)^{1/2}}{1 + e^{2\beta_S} +(3/2)e^{\beta_S}} \to \frac{1}{2}
\end{equation}
As it can be seen in Fig. 2 of the paper, this limit is already approached in a wide region for moderately negative values of $\beta_A$, and is well approximated even when $|\beta_A|< \beta_S$. Needless to say, given the symmetry of the problem all the calculations can be done in the same way for the case in which the roles of $\beta_A$ and $\beta_S$ are exchanged.

\section{The steady state rate of entropy production}
Here we formally derive the steady state entropy production rate, i.e. $\dot{S}_{\mathrm{irr}}$ appearing in the main text. Our analysis is based on general results from Stochastic Thermodynamics \cite{Seifert08,Crooks08,Horowitz13,Manzano15,Elouard17}. We consider a system ${\cal S}$ of Hamiltonian $H = \ket{\epsilon_i} \bra{\epsilon_i}$ and coupled to two heat baths at inverse (possibly negative) temperatures $\beta_A$ and $\beta_S$, respectively, 
via the Lindblad-type operators, $\mathcal{L}_A$ and $\mathcal{L}_S$. We define $\rho_A$ ($\rho_S$) as the equilibrium state of $\mathcal{L}_A$ ($\mathcal{L}_S$, such that $\mathcal{L}_i[\rho_i] = 0$ ($i=A,S$).
The respective equilibrium states for each bath are given as $\rho^\infty_A = \exp(-\beta_A H)$ and $\rho^\infty_S= \exp(-\beta_S H)$. As a consequence of the coupling to the baths, the system evolves towards some out of equilibrium steady state, $\rho_{SS}=\sum_i p_{SS}^i \ket{\epsilon_i}\bra{\epsilon_i}$. 

We now describe the evolution of the system in a quantum trajectory picture. We use the quantum jump unraveling, which assumes the presence of detectors ${\cal D}_A$ and ${\cal D}_S$ in each bath recording the emission or absorption of one excitation. This corresponds to $4$ Kraus operators, denoted as ${\cal K}_\epsilon^{\pm}$, where $\epsilon = A, S$ and $\pm$ stands for emission and absorption, respectively. The no-jump operator is denoted as ${\cal K}_0$, such that ${\cal K}^\dagger_0{\cal K}_0 + \sum_\epsilon {{\cal K}^\pm}^\dagger_\epsilon{{\cal K}^\pm}_\epsilon  = 1$. At each time step, one of these Kraus operators is stochastically applied to the state of the system.

The initial pure state of the system, $\ket{\epsilon(t_0)}$, is randomly drawn from $\rho_{SS}$ with probability $p[\epsilon(t_0)]$. The system then remains in a pure state whose stochastic evolution between time $t_0$ and $t_N$ can be fully reconstructed from the given time-ordered sequence of jumps, $\overrightarrow{\Sigma} = [{\cal K}(t_k)]_{k=0}^{k=N}$, where ${\cal K}(t_k)$ is drawn from one the Kraus operators. More specifically, at each time step $t_k$ the system has a probability to jump $P[{\cal K}(t_{k}) | \epsilon(t_k)] = \bra{\epsilon(t_k)}{\cal K}^\dagger(t_k) {\cal K}(t_k) \ket{\epsilon(t_k)}$. At time $t=t_N$ the system is in the normalized pure state $\ket{\epsilon(t_N)} ={P[\overrightarrow{\Sigma}|\epsilon(t_0)]}^{-1/2} \Pi_{k=0}^{N-1} {\cal K}(t_k) \ket{\sigma(t_k)}$, where  $P[\overrightarrow{\Sigma}|\epsilon(t_0)] = [\Pi_{k=0}^{N-1} \bra{\epsilon(t_k)}{\cal K}^\dagger(t_k) {\cal K}(t_k) \ket{\epsilon(t_k)}] $ is the conditional probability of the sequence of jumps and no jumps, starting from the initial state $\ket{\epsilon(t_0)}$. The probability of $\overrightarrow{\Sigma}$ verifies 

\begin{equation}
P[\overrightarrow{\Sigma}] = P[\overrightarrow{\Sigma}|\epsilon(t_0)].p[\epsilon(t_0)]
\end{equation}

By definition, the entropy produced during the trajectory $\overrightarrow{\Sigma}$ verifies
\begin{equation}
\label{ProdS}
\Delta_\text{i} S[\overrightarrow{\Sigma}] = \log \left( \frac{P[\overrightarrow{\Sigma}]}{\tilde{P}[\overleftarrow{\Sigma}]}  \right),
\end{equation}
where $\tilde{P}[\overleftarrow{\Sigma}]$ is now the probability for the reverse trajectory to take place. Note that this definition verifies the Second Law of thermodynamics, $\langle \Delta_\text{i} S[\overrightarrow{\Sigma}] \rangle_{\overrightarrow{\Sigma}} \geq 0$, as well as the Central Fluctuation Theorem $\langle \exp(- \Delta_\text{i} S[\overrightarrow{\Sigma}]) \rangle_{\overrightarrow{\Sigma}} = 1$, where the averages are taken over all trajectories with probability distribution $P[\overrightarrow{\Sigma}]$. \\

The probability of the reversed trajectory depends on the probability of its initial state $\tilde{p}[\epsilon(t_N)]$ and on the conditional probability $P[\overleftarrow{\Sigma}|\epsilon(t_N)]$, verifying $P[\overleftarrow{\Sigma}|\epsilon(t_N)] = \Pi_{k=N-1}^{0} \bra{\epsilon(t_k)}\tilde{{\cal K}}^\dagger(t_k) \tilde{{\cal K}}(t_k) \ket{\epsilon(t_k)} $. 
We have introduced the reversed jump operators \cite{Manzano15,Elouard17}, which can be written as

\begin{equation}
\begin{array}{l}
\tilde{{\cal K}}^\pm_\epsilon = \sqrt{\rho^\infty_\epsilon} {\cal K}^\mp_\epsilon  \sqrt{\rho^\infty_\epsilon}^{-1}\\
\tilde{{\cal K}}_0 = {\cal K}_0^\dagger
\end{array}
\end{equation}

Finally, the entropy production, Eq. (\ref{ProdS}), splits into two terms, $\Delta_\text{i} S[\overrightarrow{\Sigma}] = \Delta^\text{b}_\text{i} S[\overrightarrow{\Sigma}] + \Delta^\text{cond}_\text{i} S[\overrightarrow{\Sigma}]$, with
\begin{equation}
\begin{array}{l}
 \Delta^\text{b}_\text{i} S[\overrightarrow{\Sigma}] = \log(p[\epsilon(t_0)]) - \log(\tilde{p}[\epsilon(t_N)])    \\
\Delta^\text{cond}_\text{i} S[\overrightarrow{\Sigma}] = \log( P[\overrightarrow{\Sigma}|\epsilon(t_0)] / P[\overleftarrow{\Sigma}|\epsilon(t_N)] ).\\
\end{array}
\end{equation}
The first (second) one represents the boundary (conditional) term. Averaging over all trajectories with probability distribution $P[\overrightarrow{\Sigma}]$, the boundary term reduces to the change of entropy of the system: 
$\langle \Delta^\text{b}_\text{i} S[\overrightarrow{\Sigma}] \rangle_{\overrightarrow{\Sigma}} = \langle \log(p[\epsilon(t_0)]) - \log(\tilde{p}[\epsilon(t_N)] ) \rangle_{\overrightarrow{\Sigma}} = S(t_N) - S(t_0)$, where $S$ is the system's Von Neumann entropy, 
$S(t) = -\text{Tr}[\rho_t \log(\rho_t)]$. In the steady state, this mean boundary term vanishes and the average entropy production, ${S}_{\mathrm{irr}}=\langle \Delta_\text{i} S[\overrightarrow{\Sigma}] \rangle_{\overrightarrow{\Sigma}}$, is such that 
\begin{equation}
\dot{S}_{\mathrm{irr}} = \frac{\langle \Delta^\text{cond}_\text{i} S[\overrightarrow{\Sigma}] \rangle_{\overrightarrow{\Sigma}}} {t_N-t_0} \, ,
\end{equation}
where $\dot{S}_{\mathrm{irr}}= \mathrm{d}{S}_{\mathrm{irr}}/ \mathrm{d}t$ represents the steady state rate of entropy production.

\begin{figure*}[t]
\centering
\includegraphics[width=0.78\textwidth]{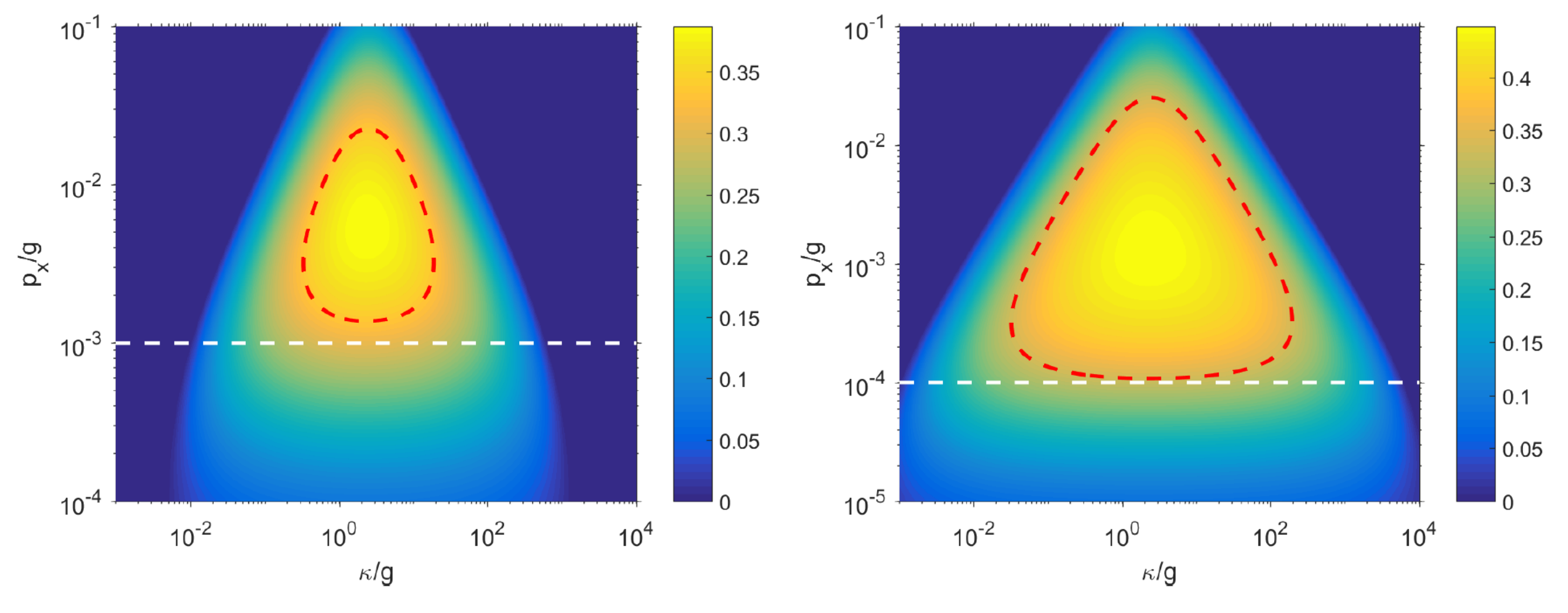} 
\vspace{-0.4cm}
\caption{Numerical results for the steady state concurrence in the real model, as a function of incoherent pumping rate $p_x/g$ and of the cavity dissipation $\kappa/g$. Panel on the left is obtained for $\gamma = 10^{-3}g$, while panel on the right for $\gamma = 10^{-4}g$. The red dashed line shows the contour line for $C_{SS} = 1/3$, while the white one shows the $p_x = \gamma$ condition, \textit{i.e.}\ the border between (infinite) positive and (infinite) negative effective hot temperature.}
\label{fig:FigS1}
\end{figure*}

\subsection{Real thermal baths}
As a paradigmatic classical case, we consider the quantum system with diamond-like level scheme (see Fig. 1a of the main text) to be coupled to two thermal baths with positive temperatures. 
It is straightforward to show that the rate $\bra{\epsilon(t_k)}{\cal K}^\dagger(t_k) {\cal K}(t_k) \ket{\epsilon(t_k)} / \bra{\epsilon(t_k)}\tilde{{\cal K}}^\dagger(t_k) \tilde{{\cal K}}(t_k) \ket{\epsilon(t_k)}$ equals $1$ only if the system has not jumped at time $t_k$. On the other hand, it reduces to $\exp(-\beta_i \delta Q_i(t_k))$, where $\beta_i = \beta_A$ ($\beta_S$), if the jump is registered in the bath coupled to $A$ ($S$), and  $\delta Q_i(t_k) = \bra{\epsilon(t_{k+1})} H \ket{\epsilon(t_{k+1})} - \bra{\epsilon(t_{k})} H \ket{\epsilon(t_{k})}$ is the variation in the system energy during the jump. In this case of real thermal baths with positive temperatures, the latter quantity actually corresponds to the heat exchanged from the system with the bath $i$. Finally, the average steady state entropy produced between $t_0$ and $t_N$ can be expressed as $\langle \Delta_\text{i} S [\overrightarrow{\Sigma}]\rangle_{\overrightarrow{\Sigma}} = - \beta_A Q_A - \beta_S Q_S$, with the average taken with respect to all trajectories and distribution $P[\overrightarrow{\Sigma}]$, and  
the steady state heat currents are defined as $\dot{Q}_i[\rho_{SS}]= \mathrm{Tr}\{ H \mathcal{L}_i (\rho_{SS}) \}$ with the Liouvillian as in the main text. 
Differentiating the former expression with respect to time straightforwardly leads to the expression used for the entropy production rate in the classical case.

Note that in this textbook situation, the average value of entropy production computed using the tools of stochastic thermodynamics exactly matches the sum of the entropy changes of the system and the baths, which constitute an isolated system as a whole. This is in full agreement with classical Thermodynamics, according to which the entropy of an isolated system can only increase. Stochastic thermodynamics allows extending the definition of entropy production to arbitrary reservoirs or ``stochastic maps'' \cite{Manzano15}. 
We now use this generalization to derive the entropy production rate in the case of effective heat baths which are allowed negative effective temperatures.

\subsection{Effective thermal baths}
In the case of effective baths $A$ and $S$ with negative inverse temperatures, $\beta_A$ and $\beta_S$, it is straightforward to see that all the used formula  remain valid. 
However, the physical interpretation of entropy production now purely relies on the tools of Stochastic Thermodynamics. In particular, the heat exchanged, $Q_A$ ($Q_S$), does not necessarily correspond to an actual energy variation of the bath $A$ ($S$). 

A concrete example to implement an effective bath can be found, e.g., in Ref. \onlinecite{EKC17}, where the absorption of a photon between $\ket{g}$ and $\ket{e}$ actually corresponds to a resonant excitation of $\ket{g}$ towards some ancillary level, $\ket{m}$, followed by the relaxation of $\ket{m}$ towards $\ket{e}$ at a very fast rate. In this specific example, the heat absorbed from the effective bath actually splits into the absorption of a photon from the drive resonant with the transition $\ket{g}\rightarrow\ket{m}$, and the spontaneous emission of a photon resonant with the transition $\ket{m}\rightarrow\ket{e}$. A negative effective temperature can be engineered, by tuning the intensity of the drive and the fast relaxation rate. The complete, microscopic description of the physical situation thus corresponds to a driven-dissipative scenario characterized by an infinite rate of entropy production because of the coupling to a zero-temperature bath. On the other hand, eliminating the ancillary level from the description brings us back to some effective thermal equilibrium where detailed balance is fulfilled and no entropy is produced. It is on this effective situation that we build our model in the present work.

\begin{figure*}[t]
\centering
\includegraphics[width=0.94\textwidth]{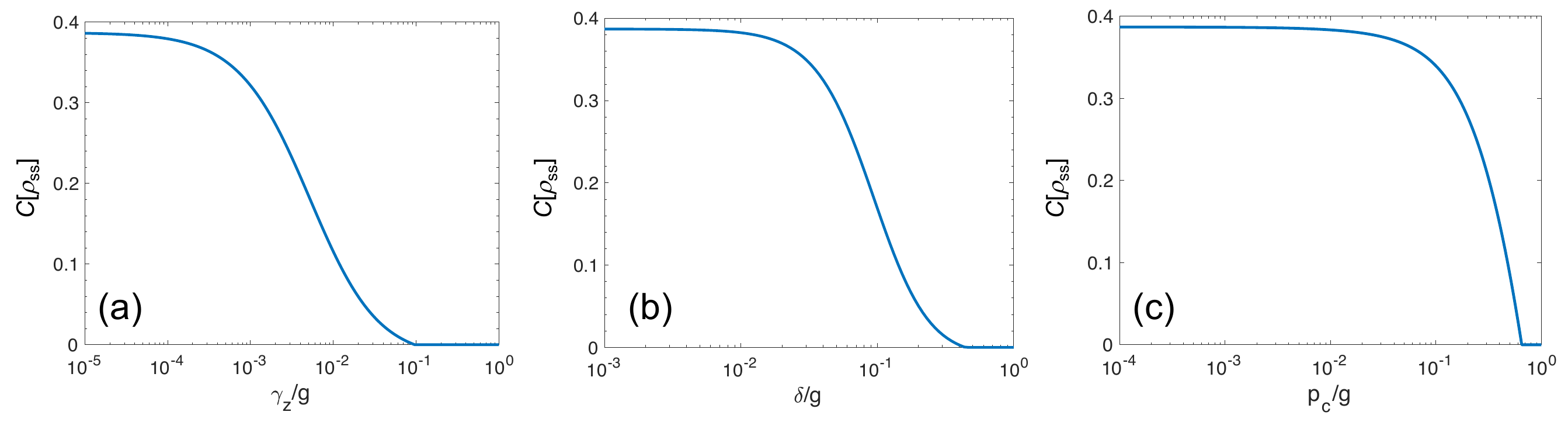}
\vspace{-0.4cm}
\caption{Steady state concurrence as a function of the main sources of decoherence and dephasing, such as (a) pure dephasing rate, (b) qubits detuning with respect to the cavity mode, and (c) finite temperature of the cavity bath.}
\label{fig:Suppl2}
\end{figure*}

\section{Steady state concurrence of the full cavity QED model}
The master equation for the full model in the main text, describing two incoherently driven quantum emitters coupled to the same cavity mode, can be most efficiently solved in steady state by 
expressing the operators on a Fock basis of occupation numbers truncated to the most suitable photon number $n_{max}$ priorly checked for convergence and then
solving numerically the equation $\mathcal{F}\rho_{SS}=0\cdot \rho_{SS}$, where $\mathcal{F}$ is the superoperator corresponding to the linear operator equation $\, i [ {\rho}, \hat{H}_0 ] +  \mathcal{L}(\rho)=0$. This is obtained by the usual mapping between the $d$-dimensional Hilbert space of the system and a $d^2$-dimensional Hilbert space via the relations
\begin{equation}
\rho = \sum_{ij} r_{ij} |i\rangle\langle j| \mapsto |\rho\rangle\rangle = \sum_{ij} r_{ij} |i\rangle |j\rangle 
\end{equation}
and
\begin{equation}
A\rho B \mapsto |A\rho B\rangle\rangle = (A \otimes B^{T})|\rho\rangle\rangle
\end{equation}
In the present case, we have $d = 2\cdot2\cdot(n_{max}+1)$ with $n_{max}\le 15$ in the simulations shown in this work, which are largely sufficient for convergence. 
The  steady state concurrence is then calculated on the reduced density matrix of the two qubits ($q_1$,$q_2$) obtained after tracing out the cavity ($C$) degrees of freedom
\begin{equation}
\rho_{q_1,q_2} = \operatorname{Tr}_{C}[\rho_{q_1,q_2,C}]
\end{equation}
and following the procedure outlined in the previous Section: this is straightforwardly implemented in a routine that applies the spin-flip $\tilde{X} = (\sigma_y \otimes \sigma_y)(X)^*(\sigma_y \otimes \sigma_y)$ operation and then finds and sorts the eigenvalues of $\rho_{q_1,q_2}\tilde{\rho}_{q_1,q_2}$.

Here we report the results of the scans over the model parameters $p$ and $\kappa$ (both in units of the qubit-cavity coupling), respectively, and for different values of the qubit relaxation rate, $\gamma$. As it can be seen in Fig.~\ref{fig:FigS1} of this Supplementary Information, for both cases there is an optimal region in the $(p_x,\kappa)$ plane where the concurrence is at a maximum. Notice that this happens for $p_x > \gamma$ and that the amount of steady state entanglement achievable in the model can exceed the $C_{SS} = 1/3$ thermal limit. In particular, for the cases that we report here, generated numerically for $n_{max} = 15$, we obtain $C_{SS}^{max} \simeq 0.3869$ for $\gamma = 10^{-3}g$ and $C_{SS}^{max} = 0.4479$ for $\gamma = 10^{-4}g$.
For $\gamma = 10^{-3}g$, the maximal concurrence is obtained at $p/ \gamma \simeq 5$ and $\kappa / g \simeq 2$, which is the case explicitly reported in Fig. 2 of the main text.


\section{Pure dephasing and inhomogeneous two-level systems}

In this last section we evaluate the robustness of the entangled steady state of the full cavity QED implementation of the model with respect to possibly detrimental processes and sources of noise in the system. In particular, we hereby analyze the performances of the modell for parameters corresponding to the maximal calculated concurrence of Fig.~\ref{fig:FigS1} for $\gamma = 10^{-3}g$, and introducing one of the following additional features:

\begin{itemize}
\item individual pure dephasing on the qubits, \textit{i.e.}\ a contribution ${\cal L}(\gamma_{i,z},\rho)= \gamma_z \left[\hat{\sigma^z_{i}}^\dagger{\rho}\hat{\sigma^z_{i}} -{\rho} \right]$ in the master equation;
\item disorder in the form of non perfectly identical qubits, or inhomogeneous size distribution in the case of artificial atoms, which we detune from the cavity in a symmetric fashion as $\omega_1 = \omega + \delta$, $\omega_2 = \omega - \delta$;
\item finite non-zero temperature of the cavity bath, by introducing an incoherent pump term ${\cal L}(p_c,\rho)= (p_c/2) \left[2\hat{a}^\dagger{\rho}\hat{a} -\hat{a} \hat{a}^\dagger {\rho} - {\rho}\hat{a} \hat{a}^\dagger \right]$.
\end{itemize}
The results are presented in Fig.~\ref{fig:Suppl2}, where it can be seen that the orders of magnitude required for noise processes to destroy the quantum coherence in the steady state are not far from those obtained in comparable situations involving a coherent pumping of the system. In particular, we notice that our device retains a still significant amount of steady state entanglement even when the pure dephasing rate equals that of the individual relaxation and pump mechanisms on the qubits, and is even more robust, albeit with a sharper transition, with respect to incoherent driving of the cavity mode. For what concerns disorder, we note that fabrication inhomogeneities are tolerated within an order of magnitude which can be qualitatively compared with the cavity-induced effective broadening $\Gamma/g \simeq g/\kappa \simeq 10^{-1}$.

\section{Practical implementation}

The theoretical cavity QED model presented in the manuscript can be practically realized in a number of possible experimental platforms, in which a steady state subradiant emission regime can be achieved. We hereby discuss two prominent examples where the physics of quantum thermal machines could be investigated in a controlled setting by using state-of-the art solid state cavity QED systems. A tolerance analysis against the main sources of decoherence and dephasing in realistic implementations, such as qubits pure dephasing and inhomogeneous broadening, as well as cavity incoherent pumping, is reported in the previous paragraph. \\
The first example relies on semiconductor quantum dots, behaving as artificial two level systems that can be coupled to a single mode of a photonic resonator. These systems allow for a controlled and fully deterministic coupling of the quantum emitters to the cavity mode \cite{Hennessy2007,Dousse2008}. Spatial control now allows to simultaneously place more than a single artificial atom in deterministic optical coupling with the same cavity mode \cite{Lyasota2015}. Either optical or electrical control of these qubits has already been demonstrated. Typical parameters for quantum dots and semiconductor microcavities made of III-V materials are $g\simeq 0.1$ meV and $\kappa$ ranging from 0.01 meV to a few meV, depending on the Q-factor of the corresponding resonator, easily allowing to access the region $\kappa / g \ge 1$ in which the thermodynamic limit of $1/3$ can be overcome, as shown in Fig. 2 of the main text. \\
As a further potential implementation of the proposed model we mention superconducting circuit quantum electrodynamic devices, in which artificial atoms are realized by Cooper pair boxes, while high-quality resonators are implemented by coplanar transmission lines \cite{Wallraff2004}. It should be noted that controlled coupling of a few qubits to a single resonator mode has already been experimentally tested \cite{DiCarlo2009,Shankar2013,Kimchi2016}. Given the high control capabilities reached for these state-of-art devices, from the tunability of the single qubits transition frequencies to their effective dissipation rates,  this platform seems particularly suited to investigate the quantum thermodynamic aspects of elementary quantum thermal machines. We also notice that the coupling rates in the range of $g\simeq 0.1$ to 10 MHz, and the dissipation rates $\kappa\simeq $ few kHz to hundreds MHz (e.g., by increasing the temperature above the superconducting critical temperature of the material constituting the transmission line), make these practical implementations of the model span almost the full available range of radiative emission properties.

\end{document}